\documentclass[runningheads]{llncs}

\usepackage{dialogue}
\usepackage{tabularx}
\usepackage{float}
\usepackage{amsmath}
\usepackage{subcaption} 
\usepackage[T1]{fontenc}
\usepackage{graphicx}
\usepackage{multirow}
\usepackage{enumitem}
\usepackage{tablefootnote}

\begin{document}
\title{Detecting LLM-Generated Short Answers and Effects on Learner Performance}
\titlerunning{Detecting LLM-Generated Short Answers}

\author{Shambhavi Bhushan\orcidID{0009-0004-3695-2334}
\and Danielle R. Thomas\orcidID{0000-0001-8196-3252}
\and Conrad Borchers\orcidID{0000-0003-3437-8979}
\and Isha Raghuvanshi\orcidID{0009-0001-9188-2285}
\and Ralph Abboud\orcidID{0000-0002-2332-0504}
\and Erin Gatz\orcidID{0000-0002-6880-5740}
\and Shivang Gupta\orcidID{0000-0002-5713-3782}
\and Kenneth R. Koedinger\orcidID{0000-0002-5850-4768}
}

\authorrunning{S. Bhushan et al.}
\institute{
Carnegie Mellon University\\
\email{\{shambhab,drthomas,cborchers,iraghuva\}@cmu.edu}\\
\email{\{shivangg,koedinger,egatz\}@andrew.cmu.edu}
\and
Learning Engineering Virtual Institute\\
\email{\{rabboud\}@levimath.org}\\
}

\maketitle 

\begin{abstract}

The increasing availability of large language models (LLMs) has raised concerns about their potential misuse in online learning. While tools for detecting LLM-generated text exist and are widely used by researchers and educators, their reliability varies. Few studies have compared the accuracy of detection methods, defined criteria to identify content generated by LLM, or evaluated the effect on learner performance from LLM misuse within learning. In this study, we define LLM-generated text within open responses as those produced by any LLM without paraphrasing or refinement, as evaluated by human coders. We then fine-tune GPT-4o to detect LLM-generated responses and assess the impact on learning from LLM misuse. We find that our fine-tuned LLM outperforms the existing AI detection tool GPTZero, achieving an accuracy of 80\% and an F1 score of 0.78, compared to GPTZero's accuracy of 70\% and macro F1 score of 0.50, demonstrating superior performance in detecting LLM-generated responses. We also find that learners suspected of LLM misuse in the open response question were more than twice as likely to correctly answer the corresponding posttest MCQ, suggesting potential misuse across both question types and indicating a bypass of the learning process. We pave the way for future work by demonstrating a structured, code-based approach to improve LLM-generated response detection and propose using auxiliary statistical indicators such as unusually high assessment scores on related tasks, readability scores, and response duration. In support of open science, we contribute data and code to support the fine-tuning of similar models for similar use cases.

\end{abstract}

\section{Introduction and Related Work}

Recent advancements in generative AI, particularly large language models (LLMs), have shown exceptional performance on tasks such as answering questions and writing \cite{hadi2023survey,Menon_Shilpa_2023}. However, their impressive capability to generate human-like text raises concerns about misuse and academic integrity, especially in online and self-paced learning environments where supervision is limited \cite{tang2024science,wu2025survey}. Existing detection tools struggle to reliably distinguish between human-authored and LLM-generated responses \cite{Chaka_2023}. In addition, little work exists on the pedagogical ramifications of LLM misuse, such as bypassing cognition or cheating in online assessments \cite{Zhai_Wibowo_Li_2024}. In contrast, misuse of instructional support in traditional technology-enhanced learning contexts, such as tutoring systems, has been reliably linked to lower learning outcomes \cite{pardos2013affective,mogessie2020confrustion}.

In response to the ongoing challenge of distinguishing LLM-generated content, researchers have developed various detection strategies and models. Previous studies highlight that ChatGPT-generated responses often follow stylized, formulaic patterns, similar to listicles or PowerPoint-style structures with brief statements supported by one to three sentences \cite{andrews2023comparing,Gao_Howard_Markov_Dyer_Ramesh_Luo_Pearson_2023}. Tools like GPTZero analyze linguistic patterns, perplexity, and burstiness to identify LLM-generated text \cite{tian2023gptzero}. However, these models, designed in controlled settings, often exhibit inconsistencies and high false positive rates in real-world applications \cite{Elkhatat_Elsaid_Almeer_2023}. An additional challenge also arises from the use of humanizing or paraphrasing tools, which can alter LLM-generated content to make it more human-like, effectively bypassing traditional detection methods \cite{aktay2024analysis,elkhatat2023evaluating}. While this remains a significant concern in high-stakes assessments, in the context of this study, low-stakes, online training with non-evaluative, self-paced micro-lessons, there is minimal incentive for participants to invest the effort or resources needed to use paraphrasing tools, as these lessons do not involve formal assessments or performance consequences.

This study proposes a structured codebook-based annotation process to systematically identify LLM-generated content in educational settings, generating human-verified training data to fine-tune LLM detection models. Unlike existing classifiers such as GPTZero, OpenAI’s AI Text Classifier, and non-causal models like RoBERTa, which struggle with short responses (less than 100 words) \cite{tian2023gptzero,Kirchner2024}, our fine-tuned model incorporates domain-specific datasets to enhance accuracy. While previous research explores model fine-tuning for specialized applications \cite{Anisuzzaman_Malins_Friedman_Attia_2024}, limited work focuses on educational contexts, particularly short-answer responses in online learning. Additionally, there is limited research on simpler, transparent techniques based on stylometry, which can serve as a crucial baseline.  \cite{Opara_2024}

Building on this approach, this study introduces several key contributions to LLM-generated text detection in educational settings. First, it provides a direct comparison between GPTZero and a fine-tuned GPT model, quantifying the performance gap and demonstrating the advantages of human-verified training data. It also compares these models against a stylometric approach open-source, based logistic regression model. Second, this study extends the detection of LLM-generated content to adult tutors in low-stakes self-paced training contexts, an underexplored domain. For purposes of this work, tutor is synonymous with learner. Third, it examines how LLM-generated responses affect learning outcomes, addressing concerns about LLM reliance fostering ``metacognitive laziness'' \cite{Fan_Tang_Le_Shen_Tan_Zhao_Shen_Li_Gašević_2024} and reducing learner engagement \cite{Zhai_Wibowo_Li_2024}. By focusing on both detection and learning effects, this study aims to enhance LLM text detection reliability while providing insights into the broader impact of AI-assisted learning. To achieve this, we address the following research questions:

\begin{itemize}[leftmargin=0pt]
\item[] \textbf{RQ1:} Can fine-tuned models be an effective solution for LLM-generated text detection of short-form responses in online lessons?
\item[] \textbf{RQ2:} How does learner performance compare between learners who have authentically produced their own responses and those who are identified of using LLM tools? 
\end{itemize}

The remainder of this paper includes Section 2 which details our methodology, including the participants and lesson context, the development of our annotation rubric for LLM-generated text, and the fine-tuning process for our GPT-4o model and the baseline models used for comparison. Section 3 presents the results, directly addressing the research questions by comparing the detection performance of the models and analyzing the differences in learning outcomes. Finally, Section 4 discusses the implications of our findings, acknowledges the study's limitations, and proposes directions for future work.

\section{Method}

\subsection{Participants, Setting and Lesson} This study analyzed responses from 534 college students employed as tutors who completed the \textit{Using Motivational Strategies} lesson. The lesson was delivered through an online platform where learners entered their responses in an open-text box or selected an MCQ option based on the condition assigned. The lesson was developed in collaboration with tutor supervisors and a university research team specializing in learning sciences and required the learners to analyze student interactions, predict the best response, and justify their reasoning.

While typical responses in such micro-lessons are relatively short, this lesson, \textit{Using Motivational Strategies lesson} exhibited a noticeable trend: responses were often longer, well-structured, and stylistically uniform, raising concerns about the use of LLMs. To investigate this, two experienced researchers manually coded 1400+ responses to identify LLM-generated versus human-authored responses, creating a labeled dataset for fine-tuning and evaluating the GPT-4o model. The dataset, codes, annotation rubric, and lesson content are located in the GitHub.\footnote{https://github.com/shambhavib20/AI-Detection}

\subsection{LLM-Generated and Human-Authored Response Rubric} To establish a reliable dataset, two researchers manually coded 1,635 responses using an annotation rubric based on learner-sourced data and prior research on LLM-generated text detection research. Responses were labeled as LLM-generated (1) or human-authored (0). Responses in which there was no consensus between the two coders were labeled as uncertain (0.5). The inter-rater reliability (IRR) was 0.64 and 0.68 for LLM-generated and human-authored responses, respectively. This moderate agreement is due to the intentional inclusion of ambiguous cases. These cases helped assess the model’s ability to handle edge cases rather than artificially inflate accuracy by limiting the dataset to clear-cut examples. 

The lessons analyzed in this study are designed to be short (5-6 minutes) and low-stakes. Since paraphrasing tools are typically used to avoid plagiarism, improve clarity, or substitute synonyms and simpler sentences \cite{chan2024}, none of which are required in these lessons, the likelihood of learners using such tools is minimal. We hypothesize that if paraphrasing tools were used, these instances would likely manifest as uncertain or edge cases as the human-authored and LLM-generated responses showed clear evidence of authorship, as elaborated in the annotation rubric. 

The lesson content involves open-ended, scenario-based learning rather than objective math problems. Basing the ground truth on human judgment allows contextual understanding of tone and intent that automated tools may miss. Research has also found that uncertainty in human decisions leads to better generalizability and error handling in models trained on human-labeled data \cite{Peterson_2019}. To tackle variation in judgments, a comprehensive annotation rubric was developed and applied consistently. Table~\ref{tab:learner-examples} displays learner-sourced examples of open responses and coding rationale to determine what constitutes LLM-generated or human-authored \cite{Fisk_2024,Herbold_2023,andrews2023comparing,Ayoub_2023}. 

\begin{table}[]
\centering
\caption{Sample size (N) and inter-rater reliability for labels, as measured in Cohen's $\kappa$.}
\setlength{\tabcolsep}{12pt} 
\renewcommand{\arraystretch}{1.3} 

\begin{tabular}{l|l|l|l}
\hline
Label & Description    & N   & IRR   \\ \hline
0     & Human-authored & 506 & 0.678 \\ \hline
0.5   & Uncertain      & 230 & - \tablefootnote{The 0.5 label has no IRR value because it was assigned to responses where raters disagreed, indicating no consistent agreement.}    \\ \hline
1     & LLM-generated  & 899 & 0.636 \\ \hline
\end{tabular}
\end{table} 

\begin{table}[htpb]
\centering
\caption{Learner-sourced examples of LLM-generated and human-authored open responses with coding rationale.}
\label{tab:learner-examples}
\renewcommand{\arraystretch}{1.3} 
\setlength{\tabcolsep}{7pt} 
\resizebox{\textwidth}{!}{%
\begin{tabular}{p{0.6\textwidth} p{0.35\textwidth}}
\hline
\textbf{Learner-Sourced Open Response Examples} & \textbf{Coding Rationale} \\
\hline
\multicolumn{2}{l}{\textbf{LLM-Generated Content}} \\
\hline
\textit{“The article emphasizes the significance of adjusting task difficulty to improve student learning and engagement. It suggests that tasks should be tailored to students' skill levels, providing clear instructions, visual aids, and personalized support… [continues for 80 more words]”} & Lengthy response that is unrelated to the lesson theme, very well-written and punctuated \\
\hline
\textit{“Positive Reinforcement: Explicitly acknowledges and praises Kevin's achievement, which is crucial for motivation. Creates a positive association… Builds Relationship: It opens a space for a casual conversation about a topic Kevin enjoys. Strengthens the teacher-student bond, fostering a more positive…”} & Lengthy and formatted response with bullets \\
\hline
\textit{“Carla, let's work together to complete the rest of your assignment. Once completed, you can explain to me how you make your scarves like we discussed earlier."”} & Inverted commas ("") at start or end of response \\
\hline
\textit{“Hey Carla, I know math can be tough...”, "Carla, I know math can be tough...", "Carla, I know math can be challenging...", "Carla, I know math can be overwhelming..."} & Similarly structured responses \\
\hline
\multicolumn{2}{l}{\textbf{Human-Authored Content}} \\
\hline
\textit{“The approach leverages Kevin's passion for baseball to create a compelling connection between his math work and his interests. Personal Connection Goal Setting Positive Reinforcement Encouragement and Support Overall, this approach ties the task to something Kevin values…”} & Improper capitalization \\
\hline
\textit{“It makes the stdnut want to work in order to receive the reward”} & Word misspelling \\
\hline
\textit{“Depending on the topic, I may change the initial practice problems we work on to fit her interests.”} & Humanized response using ‘I’ \\
\hline
\textit{“Will motivate him by encouraging that you can do this math lesson also”} & Incomplete sentences and casual tone \\
\hline
\multicolumn{2}{l}{\textbf{Uncertain}} \\
\hline
\textit{“This message connects his love for sports to his math work, making the learning process more engaging and motivating by framing it as a form of practice and improvement, just like baseball.
”} & Formal but not complex \\
\hline
\textit{"The approach I chose works best because it praises Kevin for his hard work and connects to his love for baseball. By rewarding him with a chat about baseball, it makes him feel good and keeps him motivated to do well in math."} & Use of 'I' but structured similar to typical LLM-generated responses \\
\hline
\textit{"Hey Carla, how is your day going? Did you come up with a new design for your scarves? All your previous designs are beautiful, can I look at the new one? We can work on the assignment and then you can show me your scarf ideas.
} & Structured and formal response, but conversational

\end{tabular}%
}
\end{table}

\subsection{Finetuning OpenAI’s GPT-4o} Fine-tuning was conducted via OpenAI’s API, training GPT-4o on labeled learner responses to distinguish LLM-generated and human-authored content based on patterns in structure, length, and tone. The dataset was split 80:20 for training and testing, with the split conducted at the learner level rather than the response level to avoid data leakage. Responses from 434 out of 543 students were allocated to the training set (1,302 responses), while the remaining 109 students formed the test set (333 responses).

To assess generalizability, the fine-tuned model was evaluated on two separate test sets: (1) the in-domain test set of 333 responses from the same lesson (\textit{Using Motivational Strategies}) and (2) an out-of-domain test set comprising 150 responses from six additional lessons on the platform. This out-of-domain test set on novel lessons aims at testing the generalizability of our procedure to different content in the same instructional context. For classification of these responses, the fine-tuned model was queried using a prompt aligned with the human annotation rubric. Model predictions were then compared against ground truth labels to assess effectiveness in detecting LLM-generated content.

We chose to fine-tune GPT-4o for several key reasons. First, fine-tuning allows the model to learn from the specific lesson context, helping the model distinguish authentic from LLM-generated responses within the scope of the lesson or subject area \cite{Jia_2025}. Second, by fine-tuning the model on data from similar educational contexts, meaning our learner population, it gains insights into what is expected in real student responses, whereas generic AI detectors may not be trained on such domain-specific data. Finally, fine-tuning provides flexibility: if the model makes mistakes or if additional context is required, we can adjust and retrain the model to improve its performance, which is crucial in the educational domain where accuracy is essential.

\subsection{Baseline Models for Comparison}
To evaluate the performance of our fine-tuned GPT-4o model, we compared it against two distinct models: GPTZero and a traditional open-source stylometric classifier. 

Leveraging GPTZero, a widely used commercial AI detection tool we evaluated its performance on our in-domain test set through the GPTZero API and recorded its classifications (AI, Human, or Mixed) for comparison. For transparency and a reproducible baseline, we developed a stylometric classifier using scikit-learn and writing style features based on the annotation rubric (Table 2). This writing style included a set of 13 features, including, average word length, sentence length, readability scores, punctuation, text formatting, and common human and LLM text markers. We tested the standard classifiers, Logistic Regression and Random Forest, and reported the best cross-validated performance.

\subsection{Differences in Learning Performance}
While knowing suspected LLM use is insightful, we aimed to investigate what differentiated students flagged as LLM misusers from those who were not. To examine the relationship between suspected LLM use and posttest performance, we estimated a mixed-effects logistic regression model \cite{bolker2015linear}. Each open-response predict question in the lesson was followed by a multiple-choice question (MCQ), allowing for an analysis of item correspondence. The model estimated the likelihood of answering the MCQ correctly when the corresponding open-response question was flagged as LLM-generated by our detection model. A disproportionate increase in MCQ accuracy following LLM-generated responses would suggest reliance on LLMs for both answers, raising concerns about assessment validity and learning. The model included suspected LLM use as a fixed effect and students as a random intercept to account for student-level variation in performance and repeated observations per student \cite{bolker2015linear}. Odds ratios and 95\% confidence intervals were estimated using the lme4 package in R \cite{bates_2013}, with marginal $R^2$ used to assess the variance explained by fixed effects.

\section{Results \& Discussion}
\subsection{RQ1: Can fine-tuned models effectively detect LLM-generated responses?}

\begin{figure}[htp]
    \centering
    \includegraphics[width=1\textwidth]{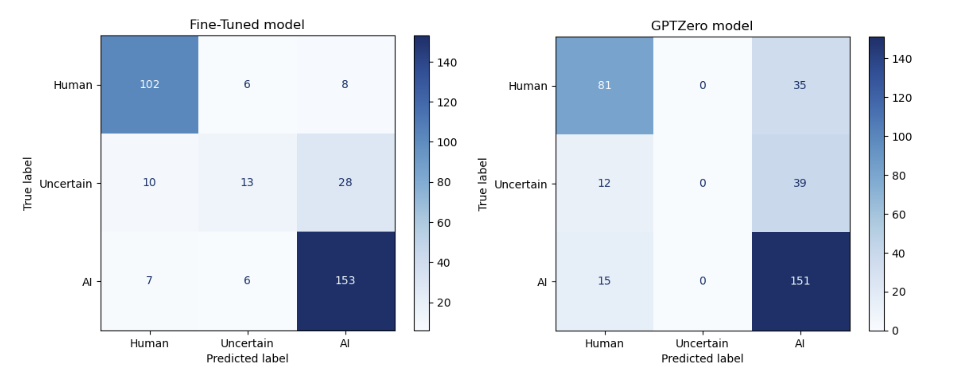}
    \caption{Confusion matrices for the fine-tuned model (left) and GPTZero (right).}
    \label{fig:propensity}
\end{figure}

The fine-tuned model demonstrated strong performance in detecting LLM-generated responses within the \textit{Using Motivational Strategies} lesson, achieving an F1 score of 86\% and an overall accuracy of 80\%, outperforming GPTZero, which had an accuracy of 70\%. The F1 average score for the fine-tuned model on the in-domain test set was 0.78 (weighted), indicating strong performance, particularly for detecting LLM-generated and human-authored responses. In comparison, GPTZero's macro average F1 score of 0.50 and weighted average of 0.64 revealed significant performance gaps, especially in handling ambiguous responses. Both models struggled with uncertain cases, but GPTZero's F1 score of 0.00 for these cases highlights its difficulty in handling edge cases, misclassifying all instances\footnote{GPTZero classifies responses as AI, Human, or Mixed. The numeric values of 1, 0, and 0.5 are then assigned to these classifications, respectively.}. Table~\ref{tab:finetuned-results} and Table~\ref{tab:gptzero-results} presents the performance metrics, while Fig.~\ref{fig:propensity}  displays the confusion matrices for both models, illustrating GPTZero’s difficulty with ambiguous cases. This highlights the limitations of general-purpose AI detectors like GPTZero which are sensitive to identifying both, LLM-generated and human-authored content with similar accuracies \cite{Kar_Bansal_Modi_Singh_2024}. An analysis of the confusion matrices for the `Uncertain' class reveals a similar pattern of overestimating LLM-generation of content. The fine-tuned model misclassified 'Uncertain' responses as LLM-generated in 28 instances and human-authored in 10 cases, while GPTZero, misclassified 86 'Uncertain' responses as LLM-generated. This suggests that both systems suspect LLM-authorship for content when the content or signals are ambiguous.

To account for generalizability across other lessons, we found that our fine-tuned model achieved an overall accuracy of 77\% and performed similarly to the out-of-domain test set. 

Comparing GPTZero and our fine-tuned model's performance against our stylometric open-source baseline, we observed that this baseline significantly outperformed GPTZero, achieving 77\% accuracy, highlighting that in AI detection of short responses, stylistic features are an accurate predictor.  

\begin{table}[h!]
\centering
\scriptsize
\caption{Performance of the fine-tuned model on in-domain and out-of-domain test sets.}
\label{tab:finetuned-results}
\renewcommand{\arraystretch}{1.2}
\resizebox{\textwidth}{!}{%
\begin{tabular}{l|llll|llll}
\hline
\multirow{2}{*}{} & \multicolumn{4}{c|}{\textbf{In-Domain Test Set}} & \multicolumn{4}{c}{\textbf{Out-of-Domain Test Set}} \\ \cline{2-9} 
                  & Precision & Recall & F1 Score & Support & Precision & Recall & F1 Score & Cases \\ \hline
Human-authored (0) & 0.86 & 0.88 & 0.87 & 116 & 0.73 & 0.92 & 0.82 & 65 \\
Uncertain (0.5)    & 0.52 & 0.25 & 0.34 & 51  & 0.25 & 0.17 & 0.20 & 18 \\
LLM-generated (1)  & 0.81 & 0.92 & 0.86 & 166 & 0.93 & 0.78 & 0.85 & 67 \\
\hline
Accuracy           &   &   & \multicolumn{1}{l}{0.80} & 333 &   &   & \multicolumn{1}{l}{0.77} & 150 \\

Macro avg.         & 0.73 & 0.69 & 0.69 & 333 & 0.64 & 0.62 & 0.62 & 150 \\
Weighted avg.      & 0.78 & 0.80 & 0.78 & 333 & 0.76 & 0.77 & 0.76 & 150 \\
\hline
\end{tabular}%
}
\end{table}

\begin{table}[h!]
\centering
\scriptsize
\caption{Performance of GPTZero on the in-domain test set.}
\label{tab:gptzero-results}
\renewcommand{\arraystretch}{1.2}
\resizebox{0.85\textwidth}{!}{%
\setlength{\tabcolsep}{7pt} 
\renewcommand{\arraystretch}{1.3} 
\begin{tabular}{l|llll}
\hline
Label               & Precision & Recall & F1 Score & Cases \\ \hline
Human-authored (0)  & 0.75 & 0.70 & 0.72 & 116 \\
Uncertain (0.5)     & 0.00 & 0.00 & 0.00 & 51 \\
LLM-generated (1)   & 0.67 & 0.91 & 0.77 & 166 \\
\hline
Accuracy            & & & \multicolumn{1}{l}{0.70} & 333 \\
Macro avg.          & 0.47 & 0.54 & 0.50 & 333 \\
Weighted avg.       & 0.60 & 0.70 & 0.64 & 333 \\
\hline
\end{tabular}%
}
\end{table}

\begin{table}[h!]
\centering
\scriptsize
\caption{Performance of open source stylometric baseline on the in-domain test set.}
\label{tab:gptzero-results}
\renewcommand{\arraystretch}{1.2}
\resizebox{0.85\textwidth}{!}{%
\setlength{\tabcolsep}{7pt} 
\renewcommand{\arraystretch}{1.3} 
\begin{tabular}{l|llll}
\hline
Label               & Precision & Recall & F1 Score & Cases \\ \hline
Human-authored (0)  & 0.79 & 0.82 & 0.81 & 116 \\
Uncertain (0.5)     & 0.47 & 0.18 & 0.26 & 51 \\
LLM-generated (1)   & 0.79 & 0.93 & 0.86 & 166 \\
\hline
Accuracy            & & & \multicolumn{1}{l}{0.77} & 333 \\
Macro avg.          & 0.69 & 0.64 & 0.64 & 333 \\
Weighted avg.       & 0.74 & 0.77 & 0.75 & 333 \\
\hline
\end{tabular}%
}
\end{table}

The improvement in accuracy achieved by the fine-tuned model over GPTZero highlights the need for human-verified training data and how general-purpose AI detectors that are verified in controlled environments are not the most accurate for specialized cases \cite{Elkhatat_Elsaid_Almeer_2023}. This discrepancy becomes evident when analyzing several misclassifications made by GPTZero. However, due to the closed-source nature of GPTZero, the reasoning behind its classifications remains unknown, making it challenging to understand the factors influencing the model's decision-making process. A few examples of these misclassifications are presented in Table~\ref{tab:output-examples}.

\begin{table}[h!]
\centering
\caption{Learner Response Classification Examples}
\label{tab:output-examples}
\renewcommand{\arraystretch}{1.3} 
\setlength{\tabcolsep}{7pt} 
\resizebox{\textwidth}{!}{%
\begin{tabular}{p{0.5\textwidth}p{0.1\textwidth}p{0.1\textwidth}p{0.1\textwidth}}
\hline
\textbf{Learner Response} & \textbf{True Label} & \textbf{GPTZero Label} & \textbf{Fine-tuned Label}  \\
\hline
\textit{"This approach will provide future extrinsic motivation for him to work hard and complete lessons earlier."} & 0 & 1 & 0  \\
\hline
\textit{"Positive Reinforcement and encouragement can boost his self-esteem and motivation to continue working hard in the future."} & 0 & 1 & 0  \\
\hline
\textit{"Collaborative Support: Working together can reduce her frustration and make the task more manageable. Encouragement: Reassuring her that she can complete the assignment boosts her confidence. Interest Connection: Linking the completion of the math assignment to your interest in scarves not only makes the task more engaging, but also more relevant to your personal interests."} & 1 & 0 & 1  \\
\hline
\textit{"Carla, let's work together to complete the rest of your assignment. You can do this, I know you can. Once we're done, I'd love to hear more about how you design your scarves and your process."} & 1 & 0 & 1  \\
\hline
\end{tabular} %
}
\end{table}

In addition, ethical concerns surrounding AI detection tools, particularly the risk of false accusations, are critical in educational settings, where misclassifications can have serious consequences \cite{Dalalah_2023}. The higher false-positive rate observed in GPTZero further highlights the limitations of generic detectors. By incorporating annotated examples specific to educational contexts, fine-tuned models can better capture domain-specific linguistic patterns, making them a more reliable tool for AI detection in online learning environments \cite{Yamtinah_Wiyarsi_Widarti_Shidiq_Ramadhani_2025}. This has direct implications for online learning, as it suggests that adopting domain-specific fine-tuning could enhance academic integrity monitoring while minimizing false accusations against students. 

\subsection{RQ2: How do posttest scores differ between LLM misusers and those who are not?}

RQ2 examined whether LLM misuse in open-ended responses influenced performance on corresponding multiple-choice questions. A mixed-effects logistic regression model was used to predict posttest scores on multiple-choice questions in relation to detected cheating on the corresponding open-ended response. The analysis revealed that learners who used LLM-generated responses in corresponding open-response questions had significantly higher odds of answering the posttest multiple-choice question correctly (OR = 2.37, p < .001). This effect size was substantial, corresponding to more than doubling ($\sim$137\% increase) in the odds of getting the multiple-choice question correct when LLM use was detected, compared to not. On the raw percent-correct scale, learners who did not use LLM had an estimated probability of 0.863 of answering correctly (95\% CI: [0.780, 0.918]), while those suspected of LLM use had a probability of 0.937 (95\% CI: [0.904, 0.960]). 

Given prior research on learning gains from scenario-based training which are considerably lower than the observed doubling of performance \cite{thomas2024tutors}, it is unlikely that this performance difference represents genuine learning. Instead the strong association between suspected LLM use and higher multiple choice scores could suggest cheating or LLM use across questions. This interpretation further aligns with Fan et al. \cite{Fan_Tang_Le_Shen_Tan_Zhao_Shen_Li_Gašević_2024}, who found that LLM tools increase learner dependence and reduce knowledge gains due to ``metacognitive laziness'' where learners rely on LLM-generated outputs rather than actively engaging in the learning process. In educational settings, particularly in self-paced training environments with limited human supervision, LLM-generated responses may allow learners to bypass meaningful engagement with instructional material, and our findings add another piece of evidence to this growing hypothesis in the field of AI use in educational settings \cite{Fan_Tang_Le_Shen_Tan_Zhao_Shen_Li_Gašević_2024}. 

\section{Limitations and Future Work}

While this study demonstrates the effectiveness of fine-tuning LLMs for detecting LLM-generated responses in short-answer online learning environments, key limitations need to be addressed in future work. The study relies on human judgment as the ground truth, which may not always be correct and introduce bias or inconsistencies \cite{Fleckenstein_Meyer_Jansen_Keller_Köller_Möller_2024}. Human annotators may vary in their interpretation of responses, especially in cases where content is ambiguous \cite{Plank_2022}, highlighting the need for further refinement in the annotation process to ensure fairness and accuracy, particularly across diverse student populations and writing styles.

Future work should also explore benchmarking the model’s performance against multiple AI detection tools and evaluating its robustness across different instructional contexts and subject areas. This will help assess its reliability in real-world educational settings. To support practical application, detection models could also be paired with deterrence strategies that discourage the use of LLMs or aid identification \cite{Shi_2024}. Such strategies include embedding white text in scenarios with specific instructions for the LLM (e.g., "make each sentence exactly seven words long"), employing audio-visual formats to prevent copying, or disabling the ability to copy and paste content altogether. These strategies not only reduce AI misuse but also make detection easier. While this study does not focus on paraphrasing tools, they remain a significant concern, and these proposed measures could help mitigate the use of these tools too. To further strengthen future models, paraphrased responses should be included in the training dataset and detection strategies should be refined to identify such content. Augmenting training data with paraphrased LLM responses would improve detection accuracy by allowing the model to learn the nuanced differences between human-authored, LLM-generated, and paraphrased content. 

Additionally, our findings show that learners suspected of LLM misuse perform better at posttest, raising concerns about genuine learning. This underscores the need for more advanced detection methods and pedagogical strategies to address AI-assisted learning, or statistical strategies to control learning gain estimates for suspected AI misuse. In particular, a promising direction for future work is to advance interventions that discourage learners from relying on LLM-generated answers that may undermine their learning. This future research may draw from interventions to mitigate students misusing scaffolding in tutoring systems (i.e., gaming the system), for instance, which have shown some success by repeating gamed exercise types \cite{baker2006adapting} or displaying learning analytics to students that spur reflection \cite{xia2020using}.

\section{Conclusion}

This study presents a structured approach for detecting LLM-generated short answers in online learning environments, demonstrating the superior performance of a fine-tuned GPT-4o model over existing, general-purpose AI detection tools such as GPTZero, and addressing both the technical challenge of accurate detection and the pedagogical implications of AI misuse in low-stakes, self-paced lessons. Our results indicate that LLM misuse is associated with substantially higher performance on posttest questions, suggesting that learners may bypass the cognitive effort required in learning by relying on AI tools. These findings underscore the importance of integrating detection systems that are both accurate and attuned to the educational context, minimizing the risk of false positives while promoting academic integrity. Institutions must integrate ethical AI education, real-time feedback, and adaptive assessments to prevent misuse while supporting meaningful learning. By aligning detection and pedagogy, AI-assisted learning can enhance rather than undermine educational integrity and effectiveness in online training.

\section*{Acknowledgments}
This work was made possible with the support of the Learning Engineering Virtual Institute. The opinions, findings and conclusions expressed in this material are those of the authors.

\bibliographystyle{splncs04}
\bibliography{references} 

\end{document}